\documentclass[11pt,lecno]{amsart}
\usepackage{indentfirst,amssymb,amsmath,amsthm}
\usepackage[mathscr]{euscript}
\usepackage[caption=false]{subfig}

\newtheorem{theorem}{Theorem}[section]

\newtheorem{lemma}{Lemma}[section]
\newtheorem{remark}{Remark}[section]
\newtheorem{corollary}{Corollary}[section]

\newcommand{\be}{\begin{equation}}
\newcommand{\ee}{\end{equation}}
\newcommand{\bea}{\begin{eqnarray}}
\newcommand{\eea}{\end{eqnarray}}
\newcommand{\beas}{\begin{eqnarray*}}
\newcommand{\eeas}{\end{eqnarray*}}

\begin{document}
\setcounter{page}{1} \setlength{\unitlength}{1mm}\baselineskip
.58cm \pagenumbering{arabic} \numberwithin{equation}{section}

\title[perfect fluid spacetimes]
{Some curvature properties of perfect fluid spacetimes }

\author[ K. De$^{*}$, U. C. De and L.S. Velimirovic ]
{ Krishnendu De $^{*}$, Uday Chand De and Ljubica S. Velimirovic}

\address
 {$^{*}$ Department of Mathematics,
 Kabi Sukanta Mahavidyalaya,
The University of Burdwan.
Bhadreswar, P.O.-Angus, Hooghly,
Pin 712221, West Bengal, India. ORCID iD: https://orcid.org/0000-0001-6520-4520}
\email{krishnendu.de@outlook.in }
\address
{ Department of Pure Mathematics, University of Calcutta, West Bengal, India. ORCID iD: https://orcid.org/0000-0002-8990-4609}
\email {uc$_{-}$de@yahoo.com}

\address
{Department of Mathematics, Faculty of Sciences and Mathematics, University of Niš, 18 000 Niš, Serbia.}
\email{ljubicavelimirovic@yahoo.com}

\footnotetext {$\bf{2020\ Math.\ Subject\ Classification\:}.$ 53B30; 53B50; 83C05; 83C56.
\\ {Key words: Perfect fluids; Robertson Walker spacetime; quasi-constant sectional curvature; Conformal curvature tensor; Static spacetime.\\
\thanks{$^{*}$ Corresponding author}
}}
\maketitle

\vspace{1cm}

\begin{abstract}
In this paper we assume that a perfect fluid is the source of the gravitational field while analyzing the solutions to the Einstein field equations. With this new and innovative approach, we investigate some curvature properties of perfect fluid spacetimes. Firstly, we provide a necessary condition for a Lorentzian manifold to be a perfect fluid spacetime. We prove that a conformally flat perfect fluid spacetime is a spacetime of quasi constant sectional curvature. Also, we acquire that a conharmonically flat perfect fluid spacetime represents the radiation era. Next, we show that if a stiff matter fluid obeys Yang’s equations, then the vorticity of the fluid vanishes. Moreover, we find that if the Ricci tensor is Killing, then the perfect fluid spacetime is (i) expansion-free and shear-free, and its flow is geodesic, however, not necessarily vorticity-free, and (ii) $\sigma$ and $p$ are constant. Besides, we establish that Ricci semi-symmetric or Ricci symmetric perfect fluid spacetimes represent either dark matter era, or phantom era. Finally, we acquire that if a perfect fluid is Ricci symmetric, then either the spacetime represents a dark matter era, or a static spacetime.

\end{abstract}

\maketitle

\section{Introduction}

The proudest glory of Applied Mathematics is general relativity (briefly, GR) theory. GR has always been recognised as the most elegant physics theory ever developed and a very challenging theory. The study of GR, which ignores quantum effects, is crucial for understanding cosmology. In GR theory, the matter content of the universe is stated by choosing the suitable energy momentum tensor. Spacetime is referred as a 4-dimensional torsion-free, time-oriented Lorentzian manifold of signature (3, 1) or, alternatively,(1, 3), as a particular subclass of semi-Riemannian manifolds. Being the natural sources of Einstein's field equations (briefly, EFEs) that are compatible with the Bianchi identities, perfect fluids perform a crucial role in the theory of GR. In numerous areas of physics, including nuclear physics, plasma physics, and astrophysics, relativistic perfect fluid  models are of immense interest. In GR theory, hypothetical distributions of matter, such as the inside of a star or, an isotropic universe, are frequently modelled using perfect fluids.\par

In a spacetime $\mathcal{M}^{n}$, for a smooth function $\psi>0$ (also called scale factor, or warping function), if $\mathcal{M}=-I \times_\psi,\mathrm{M}$, in which $I \subset\mathbb{R}$, $\mathrm{M}^{n-1}$ denotes the Riemannian manifold, then $\mathcal{M}$ is named a generalized Robertson Walker (briefly, GRW) spacetime \cite{alias1}. This spacetime represents a Robertson-Walker (briefly, RW) spacetime if the dimension of $\mathcal {M}$ is three and is of constant sectional curvature,.\par

Due to the absence of a stress tensor and heat conduction terms corresponding to viscosity, the fluid is referred to as perfect and the energy momentum tensor $T_{hk}$ is written by
\begin{equation}
\label{a9}
T_{hk}=(\sigma+p)u_{h}u_{k}+p g_{hk},
\end{equation}
where $g$ is the Lorentzian metric and $p$ and $\sigma$ denote the perfect fluid 's isotropic pressure and energy density, respectively \cite{o'neill}.  In the last equation, the velocity vector is defined by $g_{hk}u^{h}u^{k}=-1$ and $u_{h}=g_{hk}u^{k}$.\par

For a gravitational constant $\kappa$, the EFEs without a cosmological constant is described by
\begin{equation}
\label{a10}
R_{hk}-\frac{R}{2}g_{hk}=\kappa T_{hk},
\end{equation}
where $R^{l}_{hki}$ is the curvature tensor of type (1,3), $R_{hk}=R^{i}_{hki}$ and $R=g^{hk}R_{hk}$ indicate the Ricci tensor and the Ricci scalar, respectively.\par

A spacetime $\mathcal{M}$ is called a perfect fluid spacetime if the non-vanishing Ricci tensor $R_{hk}$ obeys
\begin{equation}
\label{a11}
R_{hk}=\alpha g_{hk}+\beta u_{h} u_{k},
\end{equation}
where $\alpha$ and $\beta$ are smooth functions. The foregoing equation is obtained from the equations (\ref{a9}) and (\ref{a10}) (see, \cite{manticamolinaride}).\par

Combining the equations (\ref{a9}), (\ref{a10}) and (\ref{a11}), we acquire
\begin{equation}
\label{a12}
\beta=k (p+\sigma), \,\, \alpha=\frac{k (p-\sigma)}{2-n}.
\end{equation}

Additionally, an equation of state (briefly, EOS) with the form $p = p(\sigma)$ connects $p$ and $\sigma$, and the perfect fluid spacetime is known as isentropic. Furthermore, if $p = \sigma$, the perfect fluid spacetime is referred to as stiff matter \cite{ch1}. The perfect fluid spacetime is called the dark matter era if $p+\sigma=0 $, the dust matter fluid if $p = 0$, and the radiation era if $p =\frac{\sigma}{3}$ \cite{ch1}. The universe is represented as accelerating phase when $\frac{p}{\sigma}< {-\frac{1}{3}}$. It covers the quintessence phase if $-1< \frac{p}{\sigma}< 0$ and phantom era if $\frac{p}{\sigma}< -1$.\par

The conformal and conharmonic curvature tensor denoted by $C$ and $H$ respectively, are provided in local coordinates by:
\begin{eqnarray}\label{conf}
  C_{hijk} &=&  R_{hijk}-\frac{1}{n-2}(g_{hk}R_{ij}-g_{hj}R_{ik}+g_{ij}R_{hk}-g_{ik}R_{hj})\nonumber\\&&
  +\frac{R}{(n-1)(n-2)}\{g_{hk}g_{ij}-g_{hj}g_{ik}\}
\end{eqnarray}
and
\begin{eqnarray}\label{conh}
  H_{hijk} &=&  R_{hijk}-\frac{1}{n-2}(g_{hk}R_{ij}-g_{hj}R_{ik}+g_{ij}R_{hk}-g_{ik}R_{hj}),
\end{eqnarray}
where $R_{hijk}$ denotes the curvature tensor of type (0,4).\par

If the metric of a Lorentzian manifold obeys the relation
\begin{equation}
\label{a13}
\pounds_{v}g_{hk}+2R_{hk}+2\lambda g_{hk}=0,
\end{equation}

then it is called a Ricci soliton \cite{rsh1}, where $\pounds_{v}$ is the Lie derivative operator and $\lambda$ denotes a real constant. Here, $v$ is called the potential vector field of the solitons. The solitons are known as almost Ricci solitons if $\lambda$ is a function \cite{pigola}.\par

For a non vanishing 1-form $\omega_{k}$ and a scalar function $\phi$ if the relation $\nabla_{k}u_{h} = \omega_{k}u_{h} + \phi g_{kh}$ holds, then the vector field $u$ is called torse-forming. This notion was introduced by Yano \cite{yano1} on a Riemannian manifold. It is noted that the foregoing torse-forming condition becomes $\nabla_{k}u_{h} = \phi(u_{k}u_{h}+ g_{kh})$, for a unit time-like vector.\par

To investigate a conformally flat hypersurfaces of a Euclidean space the authors \cite{chy} obtain the ensuing expression of the curvature tensor
\begin{eqnarray}\label{a1}
  R_{hijk}&=& \gamma(g_{hk}g_{ij}-g_{hj}g_{ik})\nonumber\\&&
  +\mu\{g_{hk}u_{i}u_{j}+g_{ij}u_{h}u_{k}-g_{hj}u_{i}u_{k}-g_{ik}u_{h}u_{j}\},
\end{eqnarray}
where $u_{i}$ is a unit vector, called the generator and $\gamma, \mu$ are scalars. An n-dimensional conformally flat space obeying (\ref{a1}) is named a space of quasi-constant sectional curvature and denoted by $(QC)_{n}$. However, if the equation (\ref{a1}) of the curvature tensor holds, then it can be easily verified that the space is conformally flat. So in the definition conformally flatness is not required. A Lorentzian manifold is said to be a spacetime of quasi constant sectional curvature if $u_{i}$ is a unit timelike vector. \par

The tensor $D_{hk}$ is called Killing \cite{walker} if it obeys the following condition
\begin{equation}\nonumber
  \nabla_{l}D_{hk}+\nabla_{k}D_{lh}+\nabla_{h}D_{kl}=0.
\end{equation}

Hall described Ricci recurrent ($\nabla_{l}R_{hk}=A_{l}R_{hk}$, $A_{l}$ is a covariant vector.) spacetimes in \cite{hal}. The Ricci semi-symmetry is well known to be weaker than the Ricci recurrent spacetime. In this article, we are interested in looking into the Ricci semi-symmetric perfect fluid spacetimes.\par

A spacetime is said to be semi-symmetric \cite{sz} if it obeys the relation
\begin{equation}\label{a6}
 \nabla_{l}\nabla_{m} R^{h}_{ijk}-\nabla_{m}\nabla_{l}R^{h}_{ijk}=0,
\end{equation}
where $\nabla$ indicates the covariant differentiation. Semi-symmetric spacetimes have been considered in \cite{has}. It is to be noted that the class of locally symmetric spaces ($\nabla_{l} R^{h}_{ijk}=0$) due to Cartan is a proper subset of semi-symmetric spaces.\par

A spacetime is called Ricci semi-symmetric \cite{mi} if it satisfies the relation
\begin{equation}\label{a7}
  \nabla_{l}\nabla_{m}R_{ij}-\nabla_{m}\nabla_{l}R_{ij}=0.
\end{equation}

If a Lorentzian manifold admits a timelike Killing vector field $\rho$, it is referred to as a stationary spacetime and static (\cite{sanches}, \cite{ste}, p. 283) if, additionally, $\rho$ is irrotational. We will refer to $\rho$ in this context as the static vector field, where it is assumed that spacetime is time-oriented. The product $\mathbb{R}\times S$ is called a static spacetime if it is equipped with the metric
\begin{equation}\label{sta}
  g[(t, y)]=-\beta(y)dt^2 + g_{S}[y],
\end{equation}
where $g_{S}$ denotes a Riemannian metric on S. Any static spacetime behaves like a standard one locally, with $\rho$ identifiable to $\partial t$. A spherically symmetric vacuum solution is necessarily static, according to Birkhoff's theorem \cite{bri}.\par

In \cite{o'neill}, O'Neill established that each and every RW-spacetime is a perfect fluid spacetime. Also, any GRW-spacetime of dimension four is a perfect fluid spacetime if it is a RW-spacetime and vice-versa \cite{gtt}. In \cite{she}, Shepley and Taub investigated a perfect fluid spacetime of dimension four with harmonic Weyl curvature tensor ($\nabla_{h}C^{h}_{ijk}=0$, in which $\nabla$ denotes the covariant differentiation) and the state equation $p = p(\sigma)$, and they showed that the flow is geodesic, shear-free, and irrotational. Also, the spacetime is conformally flat ($C_{hijk}=0$), and the metric is RW. Sharma investigated perfect fluid spacetimes in \cite{sha} and established that a four-dimensional perfect fluid spacetime is conformally flat if it admits a proper conformal Killing vector field ($\nabla_{h}X_{k}+\nabla_{k}X_{h}=2\rho g_{hk}$) and $\nabla_{h}C^{h}_{ijk}=0$. A perfect fluid spacetime is locally a RW-spacetime if it satisfies the EFEs with $p = p(\sigma)$, $p + \sigma \neq 0$, and a proper conformal Killing vector field which is parallel to the fluid 4-velocity \cite{col}. The existence of a concircular vector field in a conformally flat perfect fluid spacetime with closed $u_{h}$ was established by De and Ghosh in \cite{ucde}. The characteristics of perfect fluid spacetimes with Ricci solitons and Yamabe solitons were investigated by De et al. \cite{de}. The characteristics of perfect fluid spacetimes have been seen in ( \cite{kde}, \cite{kde1}, \cite{kde2}, \cite{manticamolinaride}, \cite{survey}, \cite{Mantica5}, \cite{mantica2019} ).\par

The aforementioned findings encourage us to investigate some features of four dimensional perfect fluid spacetimes. Every RW-spacetime represents a perfect fluid spacetime. Is the converse valid? Inspired by this question, in this paper, we acquire a necessary conditions when a Lorentzian manifold will be a perfect fluid spacetime.
We lay out our paper as:\par

In Section $2$,  we produce some examples of perfect fluid spacetimes. Section $3$ is concerned with the study of some curvature properties of a perfect fluid spacetime and we establish some significant theorems about radiation era, dark matter era, phantom era. Finally, we prove that a Ricci symmetric perfect fluid spacetime represents a static spacetime and the spacetime is of Petrov type I, D or O.

\section{Examples of perfect fluid spacetimes}

1. In \cite{o'neill}, O'Neill established that every RW-spacetime represents a perfect fluid spacetime. Dark matter era refers to perfect fluid spacetime with the equation of state $p+\sigma=0$ \cite{ch1}. However, so far, according to \cite{guil} a four-dimensional perfect fluid spacetime with $p+\sigma \neq 0$ is RW-spacetime if and only if it is a Yang Pure spacetime.\par

2. Any GRW-spacetime of dimension four is also a perfect fluid spacetime if the spacetime is a RW-spacetime \cite{gtt}.\par

3. Multiplying (\ref{a1}) with $g^{ij}$, we acquire
\begin{eqnarray}\label{b1}
  R_{hk} &=& \gamma(4g_{hk}-g_{hk})+\mu\{-g_{hk}+4A_{h}A_{k}-A_{h}A_{k}-A_{h}A_{k}\} \nonumber\\&&
  =(3\gamma-\mu)g_{hk}+2\mu A_{h}A_{k},
\end{eqnarray}
which represents a perfect fluid spacetime. Hence, a spacetime of quasi constant sectional curvature is a perfect fluid spacetime.\par

4. In \cite{manticamolinaride}, Mantica et al established that with the condition $C^{m}_{jkl, m}=0$, GRW-spacetimes represent perfect fluid spacetimes.\par

5. A Riemannian or a semi-Riemannian manifold is called pseudo-symmetric \cite{Chaki} if $R_{hijk}$, the components of the curvature tensor satisfies
\begin{equation}
\label{pseu1}
\nabla_{l}R_{mhij}=2v_{l} R_{mhij}+v_{m}R_{lhij}+v_{h}R_{mlik}+v_{i}R_{mhlk}+v_{k}R_{mhil},
\end{equation}
where $v_{l}$ is a non-zero covector.\par
Zhao et al \cite{kdez} proved that every pseudo-symmetric GRW-spacetime is a perfect fluid spacetime.

\section{Proof of the Theorems}
In this section, we have proved a series of theorems regarding perfect fluid spacetimes.\par

At first, we consider a Lorentzian manifold obeying an almost Ricci solitons and establish the subsequent result:
\begin{theorem}\label{th1a}
If a Lorentzian manifold admits an almost Ricci soliton whose potential vector field is a unit timelike torse-forming vector field, then it becomes a perfect fluid spacetime.
\end{theorem}
{\bf Proof :} Suppose that a Lorentzian manifold admits an almost Ricci soliton. Therefore the equation (\ref{a13}) reveals
\begin{equation}
\label{rr1}\nabla_{h}u_{k}+\nabla_{k}u_{h}+2R_{hk}+2\lambda g_{hk}=0.
\end{equation}
If $u$ is a unit timelike torse-forming vector field, then we infer $\nabla_{k}u_{h} = \phi(u_{k}u_{h}+ g_{kh})$. Using this result in (\ref{rr1}), we obtain
\begin{equation}
\nonumber
\phi(u_{k}u_{h}+ g_{kh})+\phi(u_{h}u_{k}+ g_{hk})+2R_{hk}+2\lambda g_{hk}=0,
\end{equation}
which implies
\begin{equation}
\label{r2}
R_{hk}=-(\phi+\lambda) g_{hk}-\phi u_{h}u_{k}.
\end{equation}
Thus, the spacetime becomes a perfect fluid spacetime .\par
Thus, the proof is finished.\par

A spacetime of quasi constant sectional curvature is a perfect fluid spacetime (proof is given in Example 3). Is the converse true? In this article, we establish that the converse is not true, in general. Also, we know that each and every RW-spacetime represents a perfect fluid spacetime. Is the converse valid? In this article, we show that the converse is usually not valid and demonstrate the following result:
\begin{theorem}\label{th2}
A conformally flat perfect fluid spacetime is a spacetime of quasi constant sectional curvature and a RW-spacetime with $p+\sigma \neq 0$.
\end{theorem}

{\bf Proof :}
In a conformally flat spacetime, $R_{hijk}$ (curvature tensor) is given by
 \begin{eqnarray}\label{x1}
  R_{hijk}&=& \frac{1}{2}(g_{hk}R_{ij}-g_{hj}R_{ik}+g_{ij}R_{hk}-g_{ik}R_{hj})\nonumber\\&&
  -\frac{\mathcal{R}}{6}\{g_{hk}g_{ij}-g_{hj}g_{ik}\},
\end{eqnarray}
Let us consider a conformally flat perfect fluid spacetime. Then using (\ref{a11}) in (\ref{x1}), we infer
\begin{eqnarray}\label{x2}
  R_{hijk}&=& (\frac{R}{6}+\alpha)(g_{hk}g_{ij}-g_{hj}g_{ik})\nonumber\\&&
  +\frac{\beta}{2}\{g_{hk}u_{i}u_{j}+g_{ij}u_{h}u_{k}-g_{hj}u_{i}u_{k}-g_{ik}u_{h}u_{j}\}.
\end{eqnarray}
Hence, the spacetime represents a spacetime of quasi constant sectional curvature.\par

We know that in a conformally flat spacetime $div C=0$ (`$div$' denotes the divergence). In \cite{survey}, it is established that perfect fluid spacetimes with $p+\sigma \neq 0$  and $div\, C = 0$, represent GRW-spacetimes.\par

Again, in four dimensions, every GRW-spacetime is a perfect fluid spacetime if and only if it is a RW-spacetime \cite{gtt} and thus the spacetime becomes RW.\par
Hence, the proof is finished.\par

\begin{remark}
The generalized interior Schwarzschild solutions with zero expansion is an example of a conformally flat perfect fluid  solution \cite{ste}. Only the FRW models accept the EOS of the shape $p=p(\sigma)$.
\end{remark}
Next we consider a perfect fluid spacetime with certain restriction on conharmonic curvature tensor and establish the subsequent result:
\begin{theorem}\label{th4}
A conharmonically flat perfect fluid spacetime represents the radiation era and a perfect fluid spacetime with harmonic conharmonic curvature tensor is a RW-spacetime.
\end{theorem}

{\bf Proof:} Here, we consider a conharmonically flat perfect fluid spacetime. Then using (\ref{conh}), we acquire
\begin{eqnarray}\label{conh1}
&&  R_{hijk}=\frac{1}{2}(g_{hk}R_{ij}-g_{hj}R_{ik}+g_{ij}R_{hk}-g_{ik}R_{hj}).
\end{eqnarray}
Multiplying both sides by $g^{hk}$, we find
\begin{equation}\nonumber
 R_{ij}=\frac{1}{2}(4R_{ij}-R_{ij}+Rg_{ij}-R_{ij}),
\end{equation}
which implies
\begin{equation}\label{conh2}
 R_{ij}=\frac{1}{2}R g_{ij}+R_{ij}.
\end{equation}
From the above we conclude that $R=0$.\par
Contracting the equation (\ref{a11}) and using $R=0$, we get
\begin{equation}\label{conh3}
    4\alpha- \beta =0.
\end{equation}
Hence, using (\ref{a12}), the foregoing equations yields
\begin{equation}\label{conh4}
    \frac{p}{\sigma}=\frac{1}{3},
\end{equation}
which implies that the perfect fluid spacetime represents radiation era.\par

Now using the equation (\ref{conh}), we acquire
\begin{eqnarray}\label{conh5}
  \nabla_{h} H^{h}_{ijk} &=&  \nabla_{h} R^{h}_{ijk}-\frac{1}{2}(\delta^{h}_{k} \nabla_{h}R_{ij}-\delta^{h}_{j} \nabla_{h} R_{ik}+ \nabla_{h} R^{h}_{k}g_{ij}-\nabla_{h} R^{h}_{j}g_{ik}).
\end{eqnarray}

Since $\nabla_{h} R^{h}_{ijk}=\nabla_{k} R_{ij}-\nabla_{j} R_{ik}$, therefore

\begin{eqnarray}\label{conh6}
  \nabla_{h} H^{h}_{ijk} &=&  \nabla_{k} R_{ij}-\nabla_{j} R_{ik}-\frac{1}{2}[\nabla_{k}R_{ij}-\nabla_{j} R_{ik}+\frac{1}{2}( \nabla_{k} Rg_{ij}-\nabla_{j} R g_{ik})]\nonumber\\&&
  =\frac{1}{2}[\nabla_{k}R_{ij}-\nabla_{j} R_{ik}]-\frac{1}{4}[ \nabla_{k} Rg_{ij}-\nabla_{j} R g_{ik})],
\end{eqnarray}
Suppose the conharmonic curvature tensor is harmonic, that is, $\nabla_{h} H^{h}_{ijk}=0$. It may be mentioned that harmonicity of a tensor appears in conservation laws of physics. Then the foregoing equation yields
\begin{eqnarray}\label{conh7}
  \nabla_{k}R_{ij}-\nabla_{j} R_{ik}=\frac{1}{2}[ \nabla_{k} Rg_{ij}-\nabla_{j} R g_{ik})],
\end{eqnarray}
Multiplying both sides by $g^{ij}$ gives
\begin{eqnarray}\label{conh8}
  \nabla_{k}R-\frac{1}{2}\nabla_{k} R=\frac{3}{2} \nabla_{k} R,
\end{eqnarray}
which implies $\nabla_{k}R=0$ and hence $R=$ constant.\par
Therefore, $\nabla_{h} H^{h}_{ijk}=0$ implies $\nabla_{k} R_{ij}=\nabla_{j} R_{ik}$.\par
Conversely, for a Codazzi type of Ricci tensor, we easily acquire $\nabla_{h} H^{h}_{ijk}=0$.\par
Therefore we state:\par
\begin{lemma}
  In a spacetime the conharmonic curvature tensor is harmonic if and only if the Ricci tensor is of Codazzi type.
\end{lemma}
The divergence of the conformal curvature tensor is described by

\begin{equation}\label{divc}
  \nabla_{l}C_{hki}^{l}=\frac{1}{2}[\nabla_{k}R_{hi}-\nabla_{h}R_{ki}+\frac{1}{6}(g_{ki}\nabla_{h}R-g_{hi}\nabla_{k}R)].
\end{equation}
Let us suppose that the conharmonic curvature tensor be harmonic. Then the above Lemma implies the Ricci tensor is of Codazzi type and hence $R$ is constant. Therefore the above equation infers $\nabla_{l}C_{hki}^{l}=0$. In \cite{Mantica5}, Mantica et al established that a perfect fluid spacetime with $ R=$ constant and $\nabla_{l}C_{hki}^{l}=0$ reduces to a GRW-spacetime. Again, for $n=4$, every GRW-spacetime is a perfect fluid spacetime if and only if it is a RW-spacetime \cite{gtt} and thus the spacetime becomes RW-spacetime.\par
Thus the proof is ended.\\

In \cite{guil}, Yang Pure Space is defined as a Lorentzian manifold of dimension four whose metric tensor solves Yang’s equations:
$\nabla_{l} R_{hk}-\nabla_{k}R_{hl}=0.$
A perfect fluid spacetime (in dimension four) with $\sigma +p \neq 0$ represents a RW spacetime if and only if it is a Yang pure space, according to Guilfoyle and Nolan's proof in their paper \cite{guil}, whereas here we acquired absolutely different result.\par

\begin{theorem}\label{th5a}
If a stiff matter fluid obeys Yang’s equations, then the vorticity of the fluid vanishes.
\end{theorem}
{\bf Proof :} Let the perfect fluid spacetime obey the Yang's equations, which entails
\begin{equation}\label{1b}
 \nabla_{l} R_{hk}= \nabla_{k}R_{hl}.
\end{equation}
Differentiating (\ref{a11}) covariantly yields
\begin{equation}\label{c1}
  \nabla_{l}R_{hk}=(\nabla_{l}\alpha) g_{hk}+(\nabla_{l} \beta) u_{h}u_{k}+\beta(u_{k}\nabla_{l}u_{h}+u_{h}\nabla_{l}u_{k}).
\end{equation}
Similarly, we get
\begin{equation}\label{c2}
  \nabla_{k}R_{hl}=(\nabla_{k}\alpha) g_{hl}+(\nabla_{k} \beta) u_{h}u_{l}+\beta(u_{l}\nabla_{k}u_{h}+u_{h}\nabla_{k}u_{l}).
\end{equation}
Hence, using the foregoing equations in (\ref{1b}), we acquire
\begin{eqnarray}\label{c3}
  0&=&\nabla_{l}R_{hk}-\nabla_{k}R_{hl}\nonumber\\&&
   = (\nabla_{l}\alpha) g_{hk}+(\nabla_{l} \beta) u_{h}u_{k}+\beta(u_{k}\nabla_{l}u_{h}+u_{h}\nabla_{l}u_{k})\nonumber\\&&
  -(\nabla_{k}\alpha) g_{hl}-(\nabla_{k}\beta) u_{h}u_{l}-\beta(u_{l}\nabla_{k}u_{h}+u_{h}\nabla_{k}u_{l}).
\end{eqnarray}
Multiplying the last equation by $g^{hk}$ yields
\begin{eqnarray}\nonumber
  && 4(\nabla_{l}\alpha)-(\nabla_{l}\beta)+\beta(u^{h}\nabla_{l}u_{h}+u^{k}\nabla_{l}u_{k})\nonumber\\&&
  =(\nabla_{l}\alpha)+(\nabla_{k}\beta) u^{k}u_{l}+\beta(u_{l}\nabla_{h}u^{h}+u^{k}\nabla_{h}u_{l})\nonumber,
\end{eqnarray}
which implies
\begin{eqnarray}\label{c4}
  && 3(\nabla_{l}\alpha)-(\nabla_{l}\beta)
  =(\nabla_{k}\beta )u^{k}u_{l}+\beta(u_{l}\nabla_{h}u^{h}+u^{k}\nabla_{h}u_{l}).
\end{eqnarray}

Multiplying (\ref{c4}) by $u^{l}$, we obtain
\begin{equation}\nonumber
  [3(\nabla_{l}\alpha)-(\nabla_{l}\beta)]u^{l}
  =-(\nabla_{k}\beta) u^{k}-\beta \nabla_{h}u^{h},
\end{equation}
which implies
\begin{equation}\label{c5}
  3(\nabla_{l}\alpha) u^{l}= -\beta \nabla_{h}u^{h}.
\end{equation}
We assume the perfect fluid spacetime obeys the stiff matter fluid, that is, $p=\sigma$. Then we have $\nabla_{l}\alpha=0$, where we have used the equation \eqref{a12}.
Then the previous equation tells that either $\beta=0$, or $div u^{h}=0$.\par
If $\beta=0$, then $p+\sigma=0$ which implies $\sigma=0$, since $p=\sigma$. Hence, the fluid is vacuum. This is not a physically significant scenario bearing in mind that the universe contains matter.\par

If $div u^{h}=0$, then the velocity vector field is conservative. The nature of a conservative vector field is always irrotational, thus we conclude that the perfect fluid  has zero vorticity.\par
This ends the proof.\par

Here we discuss the effect of Killing Ricci tensor in a perfect fluid spacetime and establish the subsequent theorem:

\begin{theorem}\label{th6}
If in a perfect fluid spacetime the Ricci tensor is Killing, then the perfect fluid spacetime is (i) expansion-free and shear-free, and its flow is geodesic, however, not necessarily vorticity-free, and (ii) $\sigma$ and $p$ are constant.
\end{theorem}

To prove the theorem we first state and prove the following Lemma:
\begin{lemma}
In a spacetime obeying EFEs the Ricci tensor $R_{hk}$ is Killing if and only if the energy momentum tensor is Killing.
\end{lemma}
{\bf Proof of the Lemma:}\par
 Covariant differentiation of (\ref{a10}) reveals
\begin{equation}\label{xx1}
  \nabla_{l}R_{hk}-\frac{1}{2}\nabla_{l}R g_{hk}=\kappa \nabla_{l}T_{hk}.
\end{equation}
Using the above equation, we can easily obtain
\begin{eqnarray}\label{xx2}
  &&\nabla_{l}R_{hk}+\nabla_{k}R_{lh}+\nabla_{h}R_{kl}-
  \frac{1}{2}[\nabla_{l}R g_{hk}+\nabla_{k}R g_{lh}+\nabla_{h}R g_{kl}]\nonumber\\&&
  =\kappa [\nabla_{l}T_{hk}+\nabla_{k}T_{lh}+\nabla_{h}T_{kl}].
\end{eqnarray}
Suppose the Ricci tensor is Killing, that is,
\begin{equation}\label{xx3}
  \nabla_{l}R_{hk}+\nabla_{k}R_{lh}+\nabla_{h}R_{kl}=0.
\end{equation}
Multiplying by $g^{ij}$, we get
\begin{equation}\nonumber
  \nabla_{l}R+\nabla_{k}R^{k}_{l}+\nabla_{h}R^{h}_{l}=0,
\end{equation}
which implies
\begin{equation}\label{xx4}
  \nabla_{l}R+\frac{1}{2}\nabla_{l}R+\frac{1}{2}\nabla_{l}R=0.
\end{equation}
From the previous equation we conclude that $\nabla_{l}R=0$, that is, $R=$ constant.\par
Hence equation (\ref{xx2}) entails that $T_{hk}$ is Killing.\par
Conversely, if $T_{hk}$ is Killing, then equation (\ref{xx2}) reveals
\begin{eqnarray}\label{xx5}
  &&\nabla_{l}R_{hk}+\nabla_{k}R_{lh}+\nabla_{h}R_{kl}=
  \frac{1}{2}[\nabla_{l}R g_{hk}+\nabla_{k}R g_{lh}+\nabla_{h}R g_{kl}].
\end{eqnarray}
Multiplying by $g^{hk}$, we infer
\begin{equation}\label{xx6}
  \nabla_{l}R+\nabla_{k}R^{k}_{l}+\nabla_{h}R^{h}_{l}
  =2\nabla_{l}R+\frac{1}{2}\nabla_{l}R+\frac{1}{2}\nabla_{l}R,
\end{equation}
which implies
\begin{equation}\label{xx7}
  \nabla_{l}R+\frac{1}{2}\nabla_{l}R+\frac{1}{2}\nabla_{l}R
  =3\nabla_{l}R.
\end{equation}
From the last equation we infer that $\nabla_{l}R=0$, that is, $R=$ constant.\par
Hence, from equation (\ref{xx2}) we conclude that $R_{hk}$ is Killing.\par
{\bf Proof of the main Theorem:}

In \cite{sharma2}, Sharma and Ghosh established that in a perfect fluid spacetime if $T_{hk}$ is Killing, then the perfect fluid spacetime is (i) expansion-free and shear-free, and its flow is geodesic, however, not necessarily vorticity-free, and (ii) $\sigma$ and $p$ are constant. Now applying this result and the above Lemma, we can state that the perfect fluid spacetime is (i) expansion-free and shear-free, and its flow is geodesic, however, not necessarily vorticity-free, and (ii) $\sigma$ and $p$ are constant.\par
This accomplishes the proof.\par

Now we consider a Ricci semi-symmetric perfect fluid spacetime and establish the subsequent result:

\begin{theorem}\label{th7}
If a perfect fluid spacetime is Ricci semi-symmetric, then the spacetime represents either dark matter era, or phantom era.
\end{theorem}

{\bf Proof :} Covariant differentiation of the equation (\ref{c1}) yields
\begin{align}\label{h7}
\nabla_{l}\nabla_{m}R_{ij}= \nabla_{l}\nabla_{m}\alpha g_{ij}+\nabla_{l}\nabla_{m} \beta u_{i}u_{j}
+\beta\{\nabla_{l}\nabla_{m} u_{i}u_{j}+u_{i}\nabla_{l}\nabla_{m}u_{j}\}.
\end{align}
The foregoing equation immediately gives
\begin{align}\label{h8}
\nabla_{l}\nabla_{m}R_{ij}-\nabla_{m}\nabla_{l}R_{ij}=\beta \{ u_{j}\nabla_{l}\nabla_{m} u_{i}-u_{j}\nabla_{m}\nabla_{l}u_{i}+u_{i}\nabla_{l}\nabla_{m}u_{j}-u_{i}\nabla_{m}\nabla_{l}u_{j}\}.
\end{align}
Let the perfect fluid spacetime be Ricci semi-symmetric, that is, $\nabla_{l}\nabla_{m}R_{ij}-\nabla_{m}\nabla_{l}R_{ij}=0$. Then using Ricci identity the above equation infers that
\begin{align}\label{h9}
0=\beta\{ u_{j}u_{h}R^{h}_{ilm}+u_{i}u_{h}R^{h}_{jlm}\},
\end{align}
which implies either $\beta=0$, or $\beta \neq0.$\par
Case (i): If $\beta=0$, then we have $p+\sigma=0$. Therefore, the spacetime represents the dark matter era.\par
Case (ii): If $\beta\neq0$, then $u_{j}u_{h}R^{h}_{ilm}+u_{i}u_{h}R^{h}_{jlm}=0.$ Therefore multiplying the previous relation by $g^{lm}$ we have
\begin{align}\label{h10}
u_{j}u_{h}R^{h}_{i}+u_{i}u_{h}R^{h}_{j}=0.
\end{align}
Also, from (\ref{a11}), we acquire
\begin{align}\label{h11}
u^{j}R_{ij}=\alpha u_{i}-\beta u_{i}=(\alpha-\beta)u_{i}.
\end{align}
Using (\ref{h11}) in (\ref{h10}), we get
\begin{align}\label{h12}
(\alpha-\beta)u_{i}u_{j}=0.
\end{align}
From the previous equation we conclude that $\alpha=\beta$ which implies $p+3\sigma=0$.\par Thus $\frac{p}{\sigma}=-3<-1$. Hence the spacetime represents the phantom era.\par
This ends the proof.\par

A spacetime is called conformally semi-symmetric if it fulfills the relation
\begin{equation}\label{cons}
  \nabla_{l}\nabla_{m}C^{h}_{ijk}-\nabla_{m}\nabla_{l}C^{h}_{ijk}=0.
\end{equation}

In \cite{erik}, it is established that in dimension 4, conformally semi-symmetric spacetimes are Ricci semi-symmetric spacetimes. Hence, we state the following:

\begin{corollary}
If a perfect fluid spacetime is conformally semi-symmetric, then either the spacetime represents dark matter era, or phantom era.
\end{corollary}

Again, the class of Ricci symmetric spaces ($\nabla_{l} R_{ij}=0$) is a proper subset of Ricci semi-symmetric spaces. Every semi-symmetric space is known to be Ricci semi-symmetric, but the converse is usually not true. In a Riemannian space they are equivalent for dimension three. In \cite{ta}, it has been established that for $n\geq 3$, the above stated relations are equivalent for hypersurfaces having non negative scalar curvature in a Euclidean space $E^{n+1}$.\par

From the above studies we write the subsequent result:
\begin{corollary}\label{cor1}
A Ricci symmetric perfect fluid spacetime represents either phantom era, or dark matter era.
\end{corollary}

Here, we also consider a Ricci symmetric perfect fluid spacetime to establish a different result.
\begin{theorem}\label{th8}
If a perfect fluid is Ricci symmetric, then either the spacetime represents a dark matter era, or a static spacetime.
\end{theorem}
{\bf Proof :} For a perfect fluid spacetime, we acquire
\begin{equation}\label{r1}
  R_{hk}=\alpha g_{hk}+\beta u_{h}u_{k}.
\end{equation}
Covariant differentiation of the foregoing equation yields
\begin{equation}\label{r2}
  \nabla_{l}R_{hk}=(\nabla_{l}\alpha) g_{hk}+(\nabla_{l}\beta) u_{h}u_{k}+\beta(u_{k}\nabla_{l}u_{h}+u_{h}\nabla_{l}u_{k}).
\end{equation}
By hypothesis $\nabla_{l}R_{hk}=0$. Hence from the above we infer
\begin{equation}\label{r3}
(\nabla_{l}\alpha) g_{hk}+(\nabla_{l}\beta) u_{h}u_{k}+\beta(u_{k}\nabla_{l}u_{h}+u_{h}\nabla_{l}u_{k})=0.
\end{equation}
Multiplying by $u^{k}$ gives
\begin{equation}\label{r4}
  (\nabla_{l}\alpha) u_{h}-(\nabla_{l}\beta) u_{h}-\beta \nabla_{l}u_{h}=0,
\end{equation}
since $u^{i}\nabla_{l}u_{i}=0$.\par
Again multiplying by $u^{h}$ yields
\begin{equation}\label{r5}
 - (\nabla_{l}\alpha) +(\nabla_{l}\beta) =0.
\end{equation}
Using \eqref{r5} in \eqref{r4}, we have
\begin{equation}\label{r6}
  \beta \nabla_{l}u_{h}=0.
\end{equation}

Then the previous equation entails that either $\beta=0$, or $\nabla_{l}u_{h}=0$.\par

If $\beta=0$, then $p+\sigma=0$ which implies the spacetime represents a dark matter era.\par

If $\nabla_{l}u_{h}=0$, then for a smooth vector field $v$, we acquire
\begin{equation*}
  \pounds_{v}g_{ij}=\nabla_{i}v_{j}+\nabla_{j}v_{i},
\end{equation*}
in which $\pounds$ stands for Lie derivative. Since $\nabla_{l}u_{h}=0$, therefore $\pounds_{u}g_{ij}=0$ which infer that $u$ is Killing. Also $\nabla_{l}u_{h}=0$ implies $u_{h}$ is irrotational. Hence the spacetime becomes static \cite{ste}.\par
This ends the proof.\par

\textbf {Remark}:  Since every RW-spacetime is a perfect fluid spacetime, all the results of perfect fluid spacetimes in this paper are also true in RW-spacetime.

\section{Declarations}
\subsection{Funding }
Not applicable.
\subsection{Conflicts of interest/Competing interests}
The authors declare that they have no conflict of interest.
\subsection{Availability of data and material }
Not applicable.
\subsection{Code availability}
Not applicable.
\section*{Acknowledgment}
We would like to thank the anonymous referees and the Editor for reviewing the paper carefully and their valuable comments to improve the quality of the paper.\par


\begin{thebibliography}{00}

\bibitem{alias1}L. Alias, A. Romero and M. Sanchez, {\it Uniqueness of complete spacelike hypersurfaces of constant mean curvature in generalized Robertson-Walker spacetimes}, Gen. Relativ. Gravit. {\bf 27} (1995), 71-84.

\bibitem{ch1}P.H. Chavanis, {\it Cosmology with a stiff matter era,} Phys. Rev. D {\bf92}, 103004 (2015).

\bibitem{Chaki} M.C. Chaki,{\it On pseudo symmetric manifolds}, An. \c{S}tiin\c{t}. Univ. Al. I. Cuza Ia\c{s}i. Mat. (N.S.), {\bf 33(1)}, (1987), 53-58.

\bibitem{chy} B.Y. Chen and K. Yano, {\it Hypersurfaces of a conformally flat space,} Teensor, N.S., {\bf 26} (1972), 318-322.

\bibitem {col} A. A. Coley, {\it Fluid spacetimes admitting a conformal Killing vector parallel to the
velocity vector}, Class. Quantum Grav. {\bf 8} (1991), 955-968.

\bibitem{kde} De, K., De, U.C., Syied, A.A., Turki, N.B. and Alsaeed, S., {\it Perfect Fluid Spacetimes and Gradient Solitons}, Journal of Nonlinear Mathematical Physics, {\bf29} (2022), 843–858.
\bibitem{kde1} De, K., De, U. C., {\it Perfect fluid spacetimes obeying certain restrictions on the energy-momentum tensor},  Filomat {\bf37} :11 (2023), 3483-3492.
\bibitem{kde2}  De, K. De, U.C. and Gezer, A., {\it Perfect fluid spacetimes and $k$-almost yamabe solitons}, Turk J Math (2023) 47: 1236-1246.

\bibitem{ucde} U. C. De and S. K. Ghosh, {\it On conformally flat pseudo-symmetric spaces,} Balk. J.
of Geom. and its Appl. {\bf 5} (2000), 61-64.



\bibitem{de} U.C. De, S.K. Chaubey and S. Shenawy, {\it Perfect fluid spacetimes and Yamabe solitons}, J Math Phys. {\bf 62}, 032501 (2021); https://doi.org/10.1063/5.0033967



\bibitem{ehl} J. Ehlers and W. Kundt, {\it Gravitation: An Introduction to Current Research}, ed. L. Witten, Wiley, New York, {\bf 49} 1962.

\bibitem{erik} I. Eriksson and J.M.M. Senovilla, {\it Note on (conformally) semi-symmetric spacetimes}, arXiv:0908.3246v2


\bibitem{gtt} M. Guti$\acute e$rrez and B. Olea, {\it Global decomposition of a Lorentzian manifold as a generalized Robertson-Walker space}, Differ. Geom. Appl. {\bf27} (2009), 146-156.

\bibitem{guil} B. S. Guilfoyle and B. C. Nolan, {\it Yang's gravitational theory}, Gen. Relativ. Gravit., {\bf30} (1998), 473-495.
\bibitem{rsh1}  Hamilton, R. S.,  {\it Three-manifolds with positive Ricci curvature},  J. Differ. Geom. {\bf 17} (1982), 255-306.
\bibitem{hal}G. S. Hall, {\it Ricci recurrent spacetimes}, Phys. Lett. A {\bf56} (1976), 17-18.
\bibitem{has}S. Haesen and L. Verstraelen, {\it Classification of the pseudosymmetric space–times}, J. Math. Phys. {\bf45}, 2343 (2004); https://doi.org/10.1063/1.1745129
\bibitem{bri} S.W. Hawking and G.F.R Ellis, {\it The large scale structure of spacetime}, Cambridge University Press, Cambridge, 1973.

\bibitem{manticamolinaride} C. A. Mantica, L. G. Molinari and U. C. De, {\it A condition for a perfect fluid spacetime to be a generalized Robertson-Walker spacetime}, J. Math. Phys. {\bf57} (2) (2016), 022508.

\bibitem{survey} C. A. Mantica and L. G. Molinari, {\it Generalized Robertson Walker spacetimes-A survey}, Int. J. Geom. Meth. Mod. Phys.  {\bf14} (3) (2017), 1730001 (27 pages).

\bibitem{Mantica5}C.A. Mantica, U.C. De, Y.J. Suh and L.G. Molinari, {\it Perfect fluid spacetimes with harmonic generalized curvature tensor}, Osaka J.Math., {\bf 56}, (2019), 173-182.

\bibitem{mantica2019} C. A. Mantica, L. G. Molinari, Y. J. Suh and S. Shenawy, {\it Perfect fluid, generalized Robertson Walker spacetimes, and Grays decomposition}, J. Math. Phys. {\bf60} (2019), 052506.

\bibitem{mi} V. A. Mirzoyan, {\it Structure theorems for Riemannian Ric-semisymmetric spaces},  Russian Math. (Iz. VUZ), {\bf36} (1992), 75–83.

\bibitem{o'neill}B. O'Neill, {\it Semi-Riemannian Geometry with Applications to the Relativity}, Academic Press, New York-London, 1983.
\bibitem{pigola} S. Pigola, M. Rigoli, M. Rimoldi and A. Setti, {\it Ricci almost solitons, } Ann. Sc. Norm. Super. Pisa Cl. Sci. {\bf10} (2011), 757-799.

\bibitem{sanches} M. S\'{a}nchez,  {\it On the geometry of static spacetimes}, Nonlinear Analysis: Theory, Methods and Applications,  {\bf63} 5–7, (2005), 455-463.
https://doi.org/10.1016/j.na.2004.09.009.

\bibitem{sha} R. Sharma, {\it Proper conformal symmetries of space-times with divergence-free Weyl
tensor}, J. Math. Phys. {\bf 34} (1993), 3582-3587.

\bibitem{sharma2} R. Sharma and A. Ghosh,  {\it Perfect fluid space-times whose energy-momentum tensor is conformal Killing},  J. Math. Phys. {\bf 51} (2010), 022504.


\bibitem{sz} Z.I. Szabo, {\it Structure theorems on Riemannian spaces satisfying $R(X,Y).R=0$,} J. Diff. Geom. {\bf17} (1982), 531-582.

\bibitem{she}L. C. Shepley and A. H. Taub, {\it Spacetimes containing perfect fluids and having a
vanishing conformal divergence}, Commun. Math. Phys. {\bf5} (1967), 237-256.

\bibitem{ste} H. Stephani, D. Kramer, M. Mac-Callum, C. Hoenselaers and E. Hertl, {\it Exact Solutions of Einstein’s Field Equations}, 2nd edn. Cambridge Monographs on Mathematical Physics, Cambridge University Press, 2009.

\bibitem{ta} S. Tanno, {\it Hypersurfaces satisfying a certain condition on the Ricci tensor, } Tohoku Math. J. {\bf21} (1969), 297-303.

\bibitem{walker} M. Walker and R. Penrose, {\it On quadratic first integrals of the geodesic equations for \{22\} spacetimes}, Commun. Math. Phys., {\bf18} (1970), 265-274.

\bibitem{yano1} K. Yano,  {\it On torse forming direction in a Riemannian space},  Proc. Imp. Acad. Tokyo {\bf 20} (1944), 340-345.
\bibitem{kdez}P. Zhao , U. C. De, B. Unal  and  K. De, {\it Sufficient conditions for a pseudosymmetric spacetime to be a perfect fluid spacetime}, Int. J.Geom. Methods Mod. Phys. {\bf18} (2021), 2150217 (12 pages).

\end{thebibliography}
\end{document}